\documentclass[aps,prb,showpacs]{revtex4}
\usepackage{graphicx}

\begin{document}

\title{Effect of Klein tunneling  on conductance and shot noise in ballistic graphene}  
\author{ E.B. Sonin}

\affiliation{ Racah Institute of Physics, Hebrew University of
Jerusalem, Jerusalem 91904, Israel} 

\date{\today} 

\begin{abstract}
The conductance and the Fano factor in a graphene sheet in the ballistic regime are calculated. The electrostatic potential in the sheet is modeled by a trapezoid barrier, which allows to use the exact solution of the Dirac equation in a uniform electric field in the slope areas (the two lateral sides of the trapezoid). A special attention is devoted to asymmetry with respect to the sign of the gate voltage, which is connected with the difference between the Klein tunneling and the over-barrier reflection. The comparison of the developed theory with the experiment supports the conclusion that the Klein tunneling was revealed experimentally.

\end{abstract} 
\pacs{73.23.Ad,73.50.Td,73.63.-b}
\maketitle

\section{Introduction}

The Klein tunneling\cite{Klein} is one of the most important manifestations of the relativistic Dirac spectrum in graphene \cite{Geim}. In this process an electron crosses a gap between two bands, which is a classically forbidden area, transforming from an  electron to a hole, or vise versa. The Klein tunneling was well known in the theory of narrow-gap semiconductors under the name of interband or Landau-Zener tunneling.   In the framework  of the band theory the electron wave function for the states close to a narrow gap between two broad bands must satisfy the Dirac-like equation\cite{Keld}. This can be demonstrated within the model of nearly free electrons\cite{Kane,Zim}. So the analogy with the relativistic electrodynamics was well known and exploited in the theory of semiconductors. For example, Aronov and Pikus\cite{AP} used the pseudo-Lorentz transformation (with the Fermi velocity playing the role of the light speed) treating the effect of the magnetic field on the interband (Klein-Landau-Zener) tunneling. This method was used by Shytov {\em et al.} \cite{Shyt,Shyt1} for graphene.

\begin{figure}%[t]
\centerline{\includegraphics[width=0.7\linewidth]{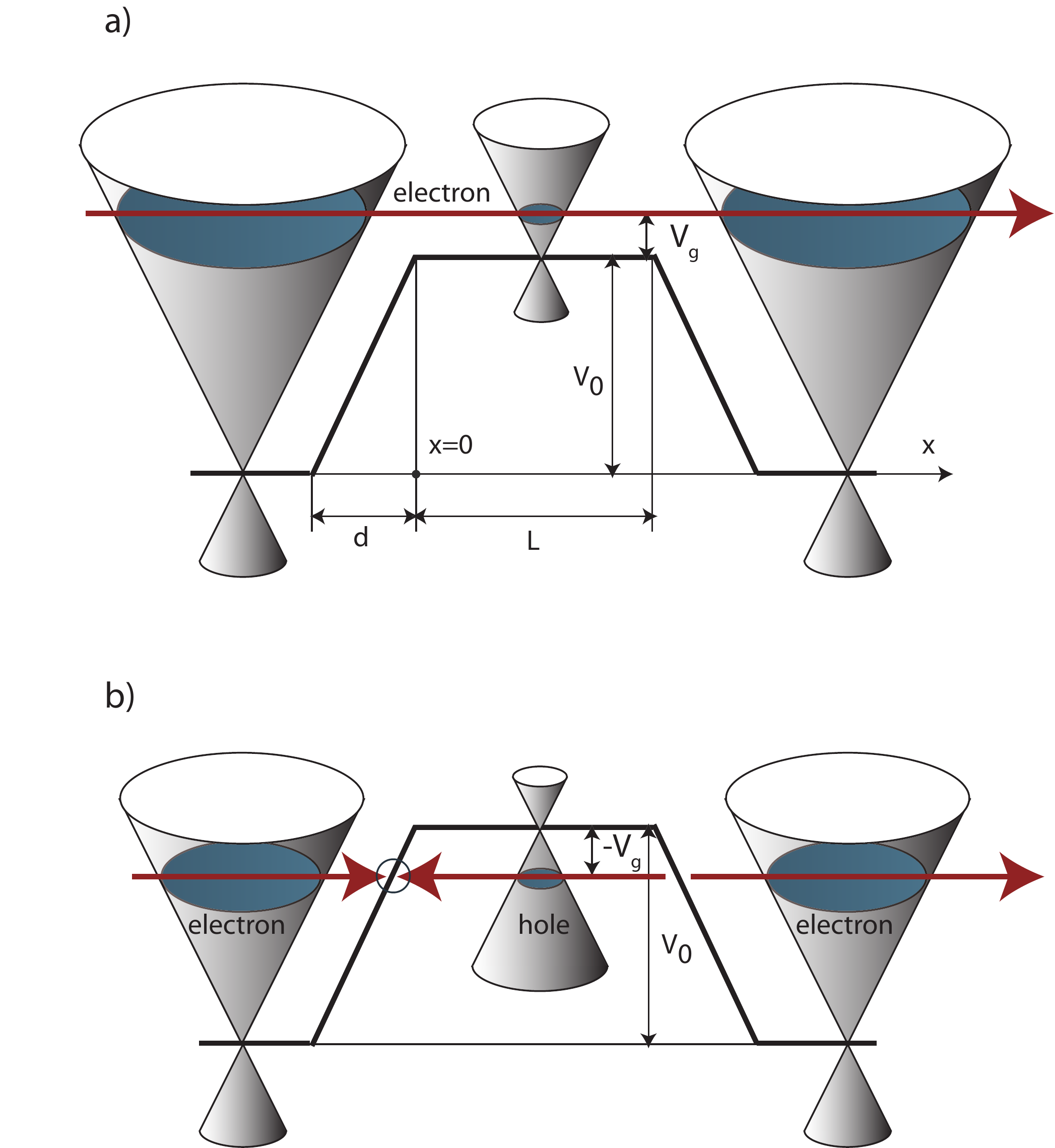}}
\caption{(color online) Trapezoidal electrostatic barrier in a graphene sheet. The thick broken solid line shows the electrostatic potential for the case, when the Fermi level in the sheet center crosses the Dirac point. (a) A positive gate voltage $V_g$ shifts the Fermi level up to the conductance band. The electron at the Fermi level is crossing the sheet from left to right without leaving the conductance band.
(b) A negative gate voltage $V_g$ shifts the Fermi level down  to the valence band. The electron crossing the sheet must tunnel from the conductance to the valence band on the left (inside the circle) and tunnel back to the conductance band on the right.  Reflected waves are not shown. } 
\label{fig1}
\end{figure}

One may expect to reveal evidence of the Klein tunneling from observations of charge transport and shot noise in a graphene sheet in the ballistic regime, which are now intensively studied experimentally\cite{Lau,Marc,Hak}.  Analyzing conductance and shot noise in a ballistic graphene sheet a commonly accepted assumption was that under electrodes the graphene is strongly doped. A further assumption, which simplified a theoretical analysis, was that the level of doping changed abruptly.  This led to  a rectangular potential barrier for electrons in a graphene sheet\cite{Kats,Been,ES}. One may expect that in reality the doping level should vary continuously.  Smooth finite-slope  potential steps  were analyzed theoretically  for $p-n$ transitions in graphene\cite{CF,Levit,Cay}. A possible model for the potential barrier might be  a trapezoid shown in Fig.~\ref{fig1}.  Recently experimental investigations of transport through tunable potential barriers were reported\cite{Gold,Gold2,Gold1}, which focused on observed asymmetry of the dependence of resistance on gate voltage with respect to the sign of voltage measured from the electrostatic potential of the Dirac point in the sheet center. In the diffusive regime asymmetry was attributed to scattering by charged impurities\cite{ch,ch1}. In the ballistic regime asymmetry is related with the Klein tunneling\cite{Gold1}.

Qualitatively the origin of asymmetry in the ballistic regime is illustrated in Fig.~\ref{fig1}.  We consider the limit of a small voltage bias, i.e., a difference of  electrochemical potentials in leads is very small.  Then only the states near the Fermi level contribute to the conductance. The small voltage bias drives electrons to the left and holes to the right.
If the gate voltage $V_g$ is positive (Fig.~\ref{fig1}a), the Fermi level crosses only the conductance band of electrons. There is no classically forbidden  zone for electrons moving above the barrier, and the transmission of the sheet is restricted only by the over-barrier reflection. 
On the other hand,  if the gate voltage $V_g$ is  negative  (Fig.~\ref{fig1}b), the Fermi level crosses the conductance band in the electrodes and the valence band in the sheet. The sheet is classically forbidden for electrons, and the charge transport is realized via holes, which annihilate with electrons moving from the left to the area of the slope of the width $d$. This is the process of the Klein tunneling. In the limit of a very steep slope (rectangular barrier) the probability of over-barrier reflection exactly coincides with the reflection from the band boundary\cite{Kats,Been}. In general these probabilities are different: the probability of over-barrier reflection decreases with the growth of positive $V_g$, while at large negative $V_g$ the Klein tunneling essentially restricts transmission and reflection probability remains finite. Therefore experimental detection of asymmetry provides evidence of the Klein tunneling\cite{Gold1}.

For studying  transport through the trapezoid barrier one should find the electron wave function inside the slope areas of the width $d$, where the electron is subject to a uniform electric field.  The Dirac equation in a uniform electric field has an exact solution in terms of confluent hypergeometric functions, which  was found long time ago by Sauter\cite{Sau}. He used it  calculating  the probability of the Klein tunneling. Kane and Blount\cite{Kane} gave an exact solution of the Dirac equation in terms of Weber  parabolic cylinder functions, which are directly connected with the confluent hypergeometric functions\cite{AS} used by Sauter\cite{Sau}.
Cheianov and  Falko\cite{CF} also found Sauter's solution but by the method valid only for not very steep slopes while the solution is exact for any slope.  Sauter's solution was also used for studying {\em p -- n} junctions in carbon nanotubes\cite{And}. This solution will be an essential component of the present analysis of conductance and short noise in graphene, though exact solutions are known also for other types potential barriers, e.g. a barrier with exponential variation of the electrostatic potential \cite{Cay}. The paper presents calculations of the conductance and the Fano factor for a trapezoid potential barrier in a ballistic graphene sheet as functions of the gate voltage.

\begin{figure}%[t]
\centerline{\includegraphics[width=0.4\linewidth]{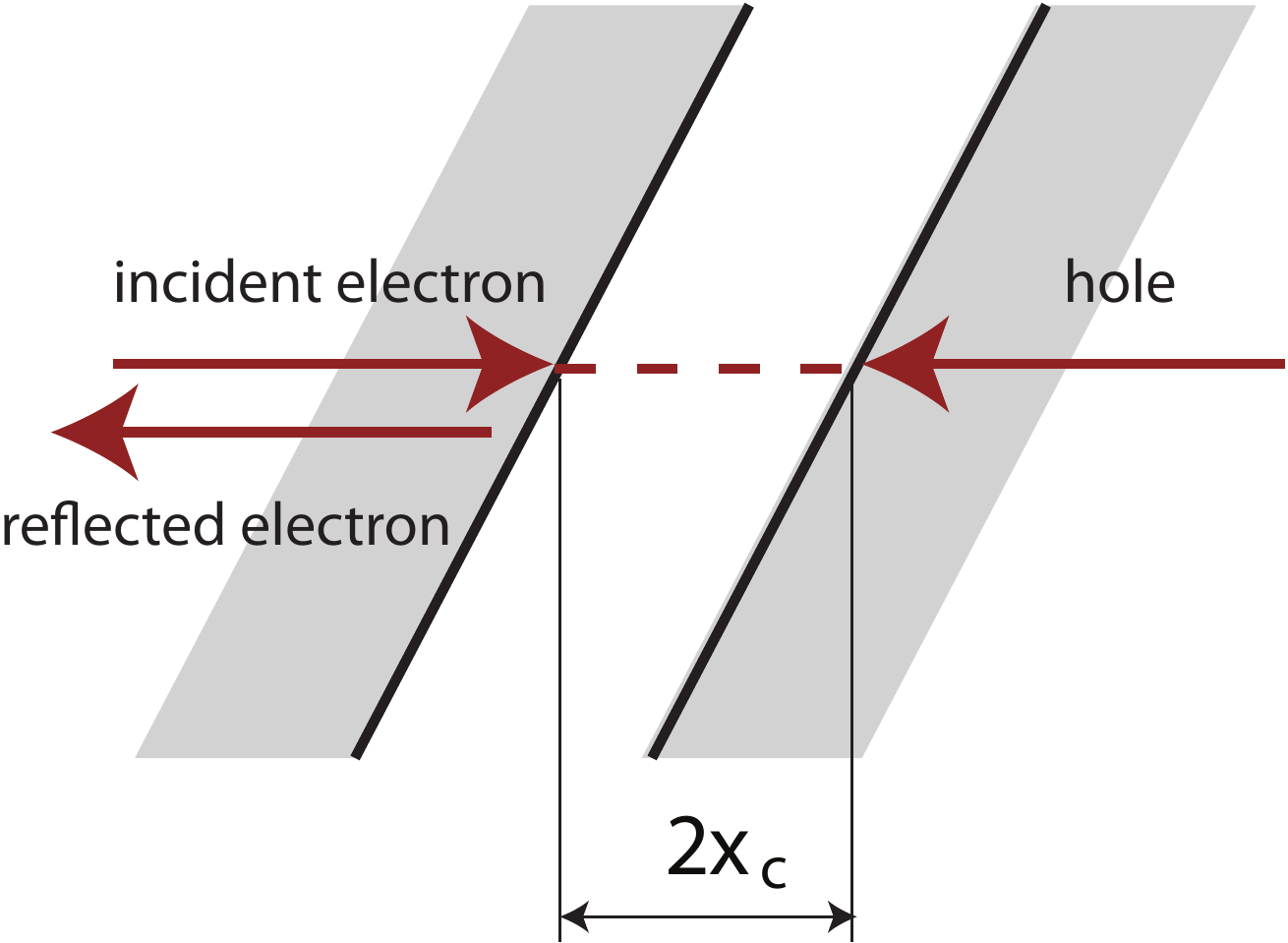}}
\caption{(color online) Klein-Landau-Zener tunneling (zoom in  on the circle in the left-slope area in Fig.~\ref{fig1}b)} 
\label{fig0}
\end{figure}

Section \ref{Dir} presents the Dirac equation for graphene and the semiclassical analysis of the Klein tunneling. Section~\ref{ex} analyzes the Klein tunneling using the known exact solution of the Dirac equation in a uniform electric field. Section~\ref{trap} studies transmission and reflection of electrons propagating across the trapezoid barrier formed in the graphene sheet by doping. Its results are used in Sec.~\ref{cond} for calculation of the conductance and the Fano factor of the sheet. The last Sec.~\ref{last} is devoted to comparison with experiment and concluding discussion.

\section{Dirac equation and semiclassical analysis} \label{Dir}

The Hamiltonian of the graphene in the presence of the electrostatic potential $V(x)$, which depends only on the coordinate $x$, is
\begin{equation}
\hat H= v_F (\hat \sigma_x \hat p_x + \hat \sigma_y \hat p_y)+ eV(x) \hat I, 
\end{equation}
where $v_F$ is the Fermi velocity, $\hat p_x=-i\hbar \partial /\partial x $ and $ \hat p_y=-i\hbar \partial /\partial y$ are components  of the momentum operator, 
$\hat I$ is a unit $2 \times 2$ matrix, and $\hat \sigma_x$, $\hat \sigma_y$, and $\hat \sigma_z$ are Pauli matrices of the pseudospin. The eigenstates are spinors
\begin{eqnarray}
{\mathbf \Psi}(x,y) =\left(\begin {array}{c} \psi_\uparrow\\ \psi_\downarrow\end{array} \right),
       \end{eqnarray}
where the components $\psi_{\uparrow \downarrow}$ are amplitudes corresponding to the eigenvalues $\pm1/2$ of the spin matrix $\sigma_z$. The components of the eigenstates with the energy $\epsilon$ satisfy the equations
\begin{eqnarray}
- \left(i{\partial \over \partial x} +{\partial \over \partial y}\right)\psi_\downarrow= {\cal K}(x)\psi_\uparrow,
\nonumber \\
-\left(i{\partial \over \partial  x} -{\partial \over \partial y}\right)\psi_\uparrow = {\cal K}(x)\psi_\downarrow,
       \end{eqnarray}
where $ {\cal K}(x) = [\epsilon -eV(x) ]/\hbar v_F$.  For the further analysis of the exact solution 
it is more convenient to perform rotation by 90$^\circ$ around the axis $y$  in the spin space introducing the spinor 
\begin{eqnarray}
{\mathbf \Psi}(x,y) =\left(\begin {array}{c} \psi_+(x) \\ \psi_-(x)\end{array} \right),
       \end{eqnarray}
where  $\psi_\pm =(\pm \psi_\uparrow + \psi_\downarrow)/\sqrt{2} $  are amplitudes corresponding to the eigenstates  of the spin matrix $\hat \sigma_x$.  After rotation the Hamiltonian becomes\cite{And}
\begin{equation}
\hat H= v_F (\hat \sigma_z \hat p_x + \hat \sigma_y \hat p_y)+ eV(x) \hat I, 
\end{equation}
and the amplitudes $\psi_\pm$ satisfy the equations
\begin{eqnarray}
-i{\partial \psi_+ \over \partial x} -{\partial \psi_- \over \partial y}  = {\cal K}(x)\psi_+,
\nonumber \\
i{\partial \psi_-\over \partial  x} +{\partial \psi_+ \over \partial y} ={\cal K}(x)\psi_-.
   \label{schX}    \end{eqnarray}

Before analyzing the exact solution of these equation for linear function ${\cal K}(x)$  (see the next section)  it is useful to  present a less accurate but physically more transparent semiclassical analysis. The semiclassical solution of Eq.~(\ref{schX}) for the $x$ dependent potential is 
\begin{eqnarray}
{\mathbf \Psi}(x,y) =\left(\begin {array}{c} \psi_+\\ \psi_-\end{array} \right)
 =\left(\begin {array}{c} {1\over 2}+{k_x+i k_y\over 2k}\\ -{1\over 2}+{k_x+i k_y\over 2k}\end{array} \right)\sqrt{k\over k_x}e^{i \int ^x k(x') dx' +ik_y y},
       \end{eqnarray}
where $k(x)=|{\cal K}(x)|$ and $k_x(x) =  \sqrt{{\cal K}(x)^2-k_y^2 }$.
The  $x$ component of the current in our presentation is
\begin{eqnarray}
j_x= ev_F{\mathbf \Psi}^\dagger \hat\sigma_z {\mathbf \Psi}= ev_F(|\psi_+|^2-|\psi_-|^2),
       \end{eqnarray}
and the spinor is normalized to the current equal to $j_x=ev_F$.

One can use the semiclassical solution for the analysis of the interband transition, which takes place for the case of negative $V_g$ (Fig.~\ref{fig1}b). Figure~\ref{fig0} shows  an electron with the wave vector ${\mathbf k}(k_x,k_y)$, which moves from the left to the right in a uniform electric field (${\cal K}=-ax$).  This is a zoom in on the Klein-Landau-Zener process inside the circle shown in the left-slope area in Fig.~\ref{fig1}b and is a slightly revised version of Fig.~114 in the book by Ziman\cite{Zim}.  At some moment the electron enters a classically  forbidden area where the $x$ component of the wave vector becomes imaginary: $k_x(x) =\pm i \sqrt{k_y^2 -{\cal K}(x)^2}$.  The point where $k(x)=0$ is the turning point of the classical trajectory. In the point where ${\cal K}(x)=0$ the electron crosses the border between states with positive and negative band energies (particle and hole bands). Let us call it ``crossing point". Choosing the crossing point to be at $x=0$, the classically forbidden area extends from $x=-x_c$ to $x_c=k_y/a$. Then according to the semiclassical approach the probability of the tunneling is
\begin{eqnarray}
T_K \sim \exp\left[-2\int_{-x_c}^{x_c}  \sqrt{k_y^2 -{\cal K}(x)^2} dx \right] =e^{-\pi k_y^2/a}.
   \label{KLZ}    \end{eqnarray}
This semiclassical analysis is not expected to provide an accurate pre-exponential factor. But remarkably the estimation fully coincides with the exact result given below.

\section{Exact solution for an uniform electric field} \label{ex}

Assuming ${\cal K}(x)=-a(x-x_0)$ the general exact solution of Eq.~(\ref{schX})  is \cite{Sau}
\begin{eqnarray}
{\mathbf \Psi}(x,y) =\left(\begin {array}{c} C_1 F(\xi, \kappa) +C_2 G(\xi, \kappa)^* \\ C_1 G(\xi, \kappa) +C_2 F(\xi, \kappa)^*\end{array} \right)e^{ ik_y y},
       \end{eqnarray}
where $\xi=(x-x_0)\sqrt{a}$, $\kappa=k_y/\sqrt{a}$, $x_0$ is the coordinate of the crossing point where ${\cal K}(x)=-a(x-x_0)=0$,
\begin{eqnarray}
F(\xi, \kappa)=e^{-i\xi^2/2}M\left(-{i\kappa^2 \over 4}, {1\over 2},i\xi^2\right),~~G(\xi, \kappa)=-\kappa \xi e^{-i\xi^2/2}M\left(1-{i\kappa^2 \over 4}, {3\over 2}, i\xi^2\right),
      \label{+field} \end{eqnarray}
and  
\begin{eqnarray}
M(a,b,z)=\sum _{n=0}^\infty{(a)_n z^n\over (b)_n n!},~~~(a)_0=1,~~~(a)_n=a(a+1)(a+2)...(a+n-1)
       \end{eqnarray}
 is the Kummer confluent hypergeometric function\cite{AS} satisfying Kummer's equation
 \begin{eqnarray}
z{d^2 f\over dz^2}+(b-z){d f\over dz} -af=0.       \end{eqnarray}
The  two constants $C_1$ and $C_2$ are determined by the boundary conditions. 
According to the known asymptotics of the Kummer functions\cite{AS},  at large distances $\xi \to \pm \infty$
\begin{eqnarray}
F(\xi, \kappa) =F_\infty(\kappa) e^{-i\xi^2/2} \xi^{i\kappa^2/2} , ~~G(\xi, \kappa)=-{\xi\over |\xi|} G_\infty( \kappa)  e^{i\xi^2/2} 
\xi^{-i\kappa^2/2},
           \end{eqnarray}
where 
\begin{eqnarray}
F_\infty ( \kappa)  =\frac{\sqrt{\pi} e^{\pi \kappa^2/8}}{\Gamma(1/2+i\kappa^2/4)}  
,~~
G_\infty( \kappa) = \frac{\sqrt{\pi}\kappa e^{\pi \kappa^2/8-i\pi /4}}{2\Gamma(1-i\kappa^2/4)} . 
           \end{eqnarray}
The functions $F_\infty ( \kappa)$ and $G_\infty ( \kappa)$ satisfy the relations\cite{AS}
\begin{eqnarray}
|F_\infty ( \kappa)|^2 ={e^{\pi \kappa^2/2}+1\over 2}
,~~
|G_\infty( \kappa)|^2 = {e^{\pi \kappa^2/2}-1\over 2}.
           \end{eqnarray}
Thus the exact solution at large distances reduces to a semiclassical solution since
 \begin{equation}
e^{i\xi^2/2}  \xi^{-i\kappa^2/2} \approx  e^{i\int_0 ^x k_x(x')dx'+ ik_y y}=e^{i\int _0^x \sqrt{a^2 x'^2 -k_y^2}dx'+ ik_y y}.
\end{equation}
One can see that very far from the crossing point ${\cal K}(x)=0$ ($\xi=0$) the electron moves parallel to the $x$ axis, and the asymptotic of the exact solution    at $\xi \to \pm \infty$ is
\begin{eqnarray}
{\mathbf \Psi}=\left(\begin {array}{c} 1 \\0 \end{array} \right)e^{-i\xi^2/2+ ik_y y}(C_1 F_\infty
\mp C_2 G_\infty^*)
\mp  \left(\begin {array}{c} 0 \\ 1\end{array} \right)e^{i\xi^2/2+ ik_y y}(C_1 G_\infty
\mp C_2 F_\infty^*).
    \label{asymp}       \end{eqnarray}

We investigate the process shown   in Fig.~\ref{fig0}: an electron moving from left either transforms after the Klein tunneling to the electron with negative energy (a hole moving from the right to the left),  or is reflected backward. Thus
 at $x \to \infty$ $(\xi \gg \kappa$), the solution should transform to the semiclassical solution
\begin{eqnarray}
{\mathbf \Psi}(x,y) = \left( \begin{array}{c} {1\over 2}+{k_x-ik_y \over 2 k} \\- {1\over 2}+{k_x-ik_y \over 2 k}\end{array}\right)\sqrt{k\over k_x}e^{-i\int_0 ^x k_x(x')dx'+ ik_y y}
\to \left( \begin{array}{c} 1 \\ 0\end{array}\right)e^{-i\xi^2/2+ ik_y y}.
     \label{spinor}       \end{eqnarray}
Using Eq.~(\ref{asymp}) and the identity $|F_\infty|^2- |G_\infty|^2=1$ one   obtains
\begin{eqnarray}
C_1 = F_\infty^* ,~~C_2=G_\infty.
           \end{eqnarray}
Then the asymptotic at $\xi \to -\infty$ is 
\begin{eqnarray}
{\mathbf \Psi}={1\over t_K}\left(\begin {array}{c} 1 \\0 \end{array} \right)e^{-i\xi^2/2+ ik_y y}
+ {r_K\over t_K} \left(\begin {array}{c} 0 \\ 1\end{array} \right)e^{i\xi^2/2+ ik_y y},
    \label{asympK}       \end{eqnarray}
where $t_K$ and $r_K$ are amplitudes of transmission and reflection determined from relations
\begin{eqnarray}
{1\over t_K}= |F_\infty|^2+ |G_\infty|^2=e^{-\pi \kappa^2/2},~~~{r_K\over t_K}=2F_\infty^*G_\infty.
           \end{eqnarray}
This yields the exact probability of the Klein tunneling $T_K =|t_K|^2=e^{-\pi \kappa^2} $, 
which coincides with the semiclassical result Eq.~(\ref{KLZ}).

\section{Transport across the trapezoid barrier} \label{trap}

\subsection{Scattering at the left side of the barrier}

Now let us consider electrons moving across the trapezoid barrier shown in Fig.~\ref{fig1}. In the area of the slope ($x<0$) electrons are in a uniform electric field, while the area $x>0$ is field-free. We look for a solution, which in the field-free area $x>0$ is a plane wave  with the current $ev_F$. For positive gate voltage $V_g$ the plane wave corresponds to an electron of positive energy [${\cal K}(x) >0$]
\begin{eqnarray}
{\mathbf \Psi}(x,y) = \left(
\begin{array}{c} {1\over 2}+{k_x+ik_y \over k}\\-{1\over 2}+{k_x+ik_y \over k}\end{array}
\right)\sqrt{k\over k_x}e^{ik_xx+ ik_y y},
     \label{spinor2}       \end{eqnarray}
while for negative $V_g$ the electron has a negative energy, and its group velocity $x$ component has a direction opposite to the  $x$ component of the wave vector ${\mathbf k}(k_x,k_y)$:
\begin{eqnarray}
{\mathbf \Psi}(x,y) = \left(
\begin{array}{c} {1\over 2}+{k_x-ik_y \over k}\\-{1\over 2}+{k_x-ik_y \over k}\end{array}
\right)\sqrt{k\over k_x}e^{-ik_xx+ ik_y y}.
     \label{spinor1}       \end{eqnarray}
The constants $C_1$ and $C_2$ in the exact solution for $x<0$ are now determined from the continuity of the spinor  at $x=0$.  Introducing the reduced gate voltage $v=eV_g/\hbar v_F \sqrt{a}$, the fitting point $x=0$ corresponds to the  argument $\xi=-v$ of the Sauter's solution, which is negative for $V_g>0$ (over-barrier reflection) and positive for $V_g<0$  (Klein tunneling).  From fitting one obtains the following expressions for $C_1$ and $C_2$ valid for the both signs of $V_g$:
\begin{eqnarray}
C_1={k+ k_x +i\mbox{sign}v\,k_y \over 2k} F^*(v,\kappa)
+ {k-k_x-i\mbox{sign}v\, k_y\over 2k} G^*(v,\kappa),
\nonumber \\
C_2= -{k-k_x-i\mbox{sign}v\, k_y\over 2k}F(v,\kappa)
- {k+k_x +i\mbox{sign}v\,k_y \over 2k}G(v,\kappa).
  \label{C1C2}     \end{eqnarray}
Here the properties $F(-\xi,\kappa)=F(\xi,\kappa)$ and $G(-\xi,\kappa)=-G(\xi,\kappa)$  were used. 

Using Eq.~(\ref{asymp})  and the calculated values of  $C_1$ and $C_2$  one obtains  the following asymptotic expression at $x \to - \infty$: 
\begin{eqnarray}
{\mathbf \Psi}={1\over t_1}\left(\begin {array}{c} 1 \\0 \end{array} \right)e^{-i\xi^2/2+ ik_y y}
+ {r_1\over t_1} \left(\begin {array}{c} 0 \\ 1\end{array} \right)e^{i\xi^2/2+ ik_y y},
    \label{asymp1}       \end{eqnarray}
where the amplitudes of the transmission, $t_1$, and of the reflection, $r_1$,  are given by
\begin{eqnarray}
{1\over t_1}= {k+k_x +i\mbox{sign}v\,k_y \over 2\sqrt{k k_x }}{\cal F}(v,\kappa)
- {k-k_x-i\mbox{sign}v\,k_y \over 2\sqrt{k k_x}}{\cal G}(v,\kappa),
        \end{eqnarray}
\begin{eqnarray}
{r_1\over t_1}= {k+k_x +i\mbox{sign}v\,k_y \over 2\sqrt{k k_x}}{\cal G}(v,\kappa)^*
- {k-k_x-i\mbox{sign}v\,k_y \over 2\sqrt{k k_x}} {\cal F}(v,\kappa)^*.
        \end{eqnarray}
Here
\begin{eqnarray}
{\cal F}(v,\kappa) =F^*(v,\kappa)F_\infty(\kappa)+G(v,\kappa) G_\infty(\kappa)^*,~~~{\cal G}(v,\kappa)= G^*(v,\kappa) F_\infty+F(v,\kappa)G_\infty(\kappa)^*.
        \end{eqnarray}
Eventually the transmission probability at the left slope of the trapezoid barrier is given by
\begin{eqnarray}
T_1=|t_1|^2  =\left[{k+k_x\over 2k_x}|{\cal F}(v,\kappa)|^2+{k-k_x\over 2k_x}|{\cal G}(v,\kappa)|^2
- \mbox{sign}v{k_y\over k_x}\mbox{Im}\{{\cal F}(v,\kappa)^*{\cal G}(v,\kappa)\}\right]^{-1}.
      \label{T1}     \end{eqnarray}

According to Fig.~\ref{fig1} the electric field is absent deep in the electrode at $x<-d$ . So in the point $x=-d$ the field solution must transform to a plane-wave solution again. However, there is no significant reflection of the electron in this area, if the asymptotic   expression Eq.~(\ref{asymp1}) is valid at $x=-d$. Indeed, according   to Eq.~(\ref{asymp1}) the electron propagates normally to the barrier and cannot be reflected. The conditions for it are $|\xi(-d)| \approx d  \sqrt{a} \gg 1$ and $|\xi(-d)| \gg \kappa$. In particular, this means that
\begin{eqnarray}
{e(V_0-V_g)\over \hbar v_F \sqrt{a}} \approx  {eV_0\over \hbar v_F \sqrt{a}} ={k_0 \over \sqrt{a}} =\sqrt{k_0 d} \gg 1,
     \label{Vg}   \end{eqnarray}
where $k_0=eV_0 /\hbar v_F \sqrt{a}$ is the modulus of the wave vector at the Fermi level inside the electrode at $x<-d$. As far as this condition is satisfied, neither $k_0$ nor $d$ affect  the results of the analysis, which depend only on their ratio $a =k_0/d$ proportional to the electric field in the area of the slope $-d<x<0$.

Let us consider various limits of the obtained expression for the transmission. The  very steep slope (rectangular barrier) corresponds to very small $|v|$ and $\kappa$. In this limit  ${\cal F}(v,\kappa) =F^*(v,\kappa)\to 1$ and ${\cal G}(v,\kappa) =G^*(v,\kappa)\to 0$,
and one obtains that transmission independent of the sign of $V_g$. i.e., the difference between 
the Klein tunneling and the over-barrier reflection vanishes: 
\begin{eqnarray}
T_1 ={2 k_x \over   k+k_x}.
            \end{eqnarray}
This result is valid for small $|V_g|$ much less than $\hbar v_F \sqrt{a} /e$ ($|v|\ll 1$).  

Let us consider now  the opposite limit of high $|V_g|$ when $|v| \gg \kappa$, and  
\begin{eqnarray}
{\cal F}(v,\kappa) \to [|F_\infty(\kappa)|^2-\mbox{sign}v\,|G_\infty(\kappa)|^2]e^{iv^2/2},~~~{\cal G}(v,\kappa)\to 2\theta(-v) F_\infty(\kappa)G_\infty(\kappa)^*e^{-iv^2/2}.
        \end{eqnarray}
According to the definitions $v$ and  $\kappa$ the  condition $|v| \gg \kappa$ means that $k \approx k_x \gg k_y$, i.e., the incident electron moves nearly normally to the barrier. Then one should keep only the first term in Eq.~(\ref{T1}):
\begin{eqnarray}
T_1= \frac{1}{[|F_\infty(\kappa)|^2-\mbox{sign}v\,|G_\infty(\kappa)|^2]^2}
= \frac{4}{[e^{\pi \kappa^2/2}+1-\mbox{sign}v\,(e^{\pi \kappa^2/2}-1)]^2}.
        \end{eqnarray}
This yields the probability of the Klein tunneling given by Eq.~(\ref{KLZ}) for negative $v$ and ideal transmission $T_1=1$ for positive $v$, i.e., for large positive $v$ the over-barrier reflection vanishes,  

For further calculation of the transmission of the whole barrier one needs also to know the parameters of the process time-reversed with respect to scattering at the left slope of the barrier. The time-reversed state corresponds to the negative current $-ev_F$ (from the right  to the left) and is described in the field-free region $x>0$ by the spinors in Eqs.~(\ref{spinor2}) and (\ref{spinor1}) after replacing $k_x$ by $-k_x$. Also the roles of an incident and a reflected wave in the asymptotic expression Eq.~(\ref{asymp1}) are interchanged, and the transmission and the reflection amplitudes are determined from those for the original process as
\begin{eqnarray}
{1\over \tilde t_1(k_x,k_y)} =   {r_1(-k_x,k_y)\over t_1(-k_x,k_y)},~~{\tilde r_1(k_x,k_y)\over \tilde t_1(k_x,k_y)}={1\over  t_1(-k_x,k_y)}.
            \end{eqnarray}

\subsection{Scattering at the right slope of the barrier}

The analysis of this process is similar to that done for scattering at the left slope: one should fit the exact solution in the region $x>L$ of the uniform electric field, which has now an opposite direction with respect to that at the left slope, i.e., ${\cal K}(x)= a(x-x_0)$. Without repeating all details we summarize here the results.

The exact solution at the right slope is complex conjugate to that at the left slope. Asymptotically  the exact solution in the field region is a wave propagating to $x \to \infty$:
\begin{eqnarray}
{\mathbf \Psi}(x,y) =  \left( \begin{array}{c} 1 \\ 0\end{array}\right)e^{i\xi^2/2+ ik_y y}.
     \label{spinor3}       \end{eqnarray}
The solution in the field area $x>L$ must fit to the solution in the field-free region $x<L$, where there is a superposition of an incident and a reflected wave, which is
\begin{eqnarray}
{\mathbf \Psi}(x,y) ={1\over t_2} \left(
\begin{array}{c} {1\over 2}+{k_x+ik_y \over k}\\-{1\over 2}+{k_x+ik_y \over k}\end{array}
\right)\sqrt{k\over k_x}e^{ik_xx+ ik_y y}
+{r_2\over t_2} \left(
\begin{array}{c} {1\over 2}+{-k_x+ik_y \over k}\\-{1\over 2}+{-k_x+ik_y \over k}\end{array}
\right)\sqrt{k\over k_x}e^{-ik_xx+ ik_y y}
     \label{sp}       \end{eqnarray}
for positive $V_g$ (over-barrier reflection) and
\begin{eqnarray}
{\mathbf \Psi}(x,y) ={1\over t_2}
 \left(
\begin{array}{c} {1\over 2}+{k_x-ik_y \over k}\\-{1\over 2}+{k_x-ik_y \over k}\end{array}
\right)\sqrt{k\over k_x}e^{-ik_xx+ ik_y y}
+{r_2\over t_2}  \left(
\begin{array}{c} {1\over 2}+{-k_x-ik_y \over k}\\-{1\over 2}+{-k_x-ik_y \over k}\end{array}
\right)\sqrt{k\over k_x}e^{ik_xx+ ik_y y}
     \label{sp1}       \end{eqnarray}
 for negative $V_g$ (Klein tunneling). The fitting at the point $x=L$ gives 
 \begin{eqnarray}
{1\over t_2}= \left[{k+k_x -i\mbox{sign}v\,k_y \over 2\sqrt{k k_x }}{\cal F}(v,\kappa)
+ {k-k_x+i\mbox{sign}v\,k_y \over 2\sqrt{k k_x}}{\cal G}(v,\kappa)\right]e^{-i \mbox{\small sign}v\, k_x L},
\nonumber \\
{r_2\over t_2}= -\left[{k+k_x +i\mbox{sign}v\,k_y \over 2\sqrt{k k_x}}{\cal G}(v,\kappa)
+ {k-k_x+i\mbox{sign}v\,k_y \over 2\sqrt{k k_x}} {\cal F}(v,\kappa)\right]e^{i \mbox{\small sign}v\, k_x L}.
    \label{sl2}    \end{eqnarray}
This yields the transmission probability $T_2=|t_2|^2$ equal to $T_1$ at the left slope of the barrier.

\subsection{Transmission through the whole barrier} 

The  transmission through the whole barrier depends on whether the electron can propagate inside the barrier coherently without losing its original phase\cite{ES} [see the phase factors $e^{\pm i \mbox{\small sign}v\, k_x L}$ in Eq.~(\ref{sl2})]. This is not the case for $L$ long enough. Then  the total transmission $T$ is combined  not from amplitudes but from probabilities, treating two slopes as uncorrelated scatters. Keeping in mind that $T_1=T_2$ one obtains\cite{Dat} 
\begin{eqnarray}
T={T_1 T_2\over 1-R_1R_2}= {T_1 \over 2-T_1},
     \label{Tuncor}       \end{eqnarray}
where $R_{1,2}=|r_{1,2}|^2 =1-T_{1,2}$ are probabilities of reflection at the left and the right slope. 

If the phase correlation takes place one should look for the solution in the electric field at $x<0$, which at $x\to -\infty$ has the same form as Eq.~(\ref{asymp1}), 
\begin{eqnarray}
{\mathbf \Psi}={1\over t}\left(\begin {array}{c} 1 \\0 \end{array} \right)e^{-i\xi^2/2+ ik_y y}
+ {r\over t} \left(\begin {array}{c} 0 \\ 1\end{array} \right)e^{i\xi^2/2+ ik_y y},
    \label{asympR}       \end{eqnarray}
but the transmission and the reflection amplitudes are determined now from fitting to the spinors Eqs.~ (\ref{sp}) or  (\ref{sp1}) in the point $x=0$. This yields the following expressions for them [compare with Eq.~(9) in Ref.~\onlinecite{Levit}]:
\begin{eqnarray}
{1\over t}={1\over t_1 t_2}+{\tilde r_1 r_2 \over \tilde  t_1 t_2}
= \left[\cos k_x L -{ik \over k_x}\sin k_x L\right] {\cal F}(v,\kappa)^2
+ \left[\cos k_x L +{ik \over k_x}\sin k_x L\right]{\cal G}(v,\kappa)^2
\nonumber \\
+{2k_y \over k_x}\sin k_x L  {\cal F}(v,\kappa){\cal G}(v,\kappa).
    \label{totA}    \end{eqnarray}
These expressions can also be used for the case, when there are no propagating modes in the field-free area $0<x<L$, i.e., $k_x$ is imaginary and corresponds to an evanescent state in the classically forbidden barrier area. In this case one should analytically continue the expressions replacing $ \sin k_x L /i k_x$  with $\sinh pL/p$, where $p=ik_x$ is real if $k_x$ is imaginary.

\section{Conductance and shot noise} \label{cond}

\subsection{Incoherent ballistic transport}

Let us consider the case of the  graphene-sheet length  of the  $L$ long enough, when the contribution of evanescent modes is not important, and there is no phase correlation between two tunneling events at the two sides of the barrier. Then the conductance does not depend on $L$ and is given by 
\begin{equation}
g={g_0 \over \sqrt{a} }  \int_0^{e|V_g|/\hbar v_F}T dk_y= g_0  \int_0^{|v|}T d\kappa . 
      \label{con} \end{equation}
where $g_0 = 4e^2 W \sqrt{a}/\pi h $, $W$ is  the width  of the graphene sheet,  and the transmission probability $T$ is determined with help of Eqs.~ (\ref{T1})  and (\ref{Tuncor}). One may replace summation over transversal components $k_y$ by integration assuming that $W$ exceeds all other spatial scales($L$ and $1/\sqrt{a}$).

\begin{figure}%[t]
\centerline{\includegraphics[width=0.7\linewidth]{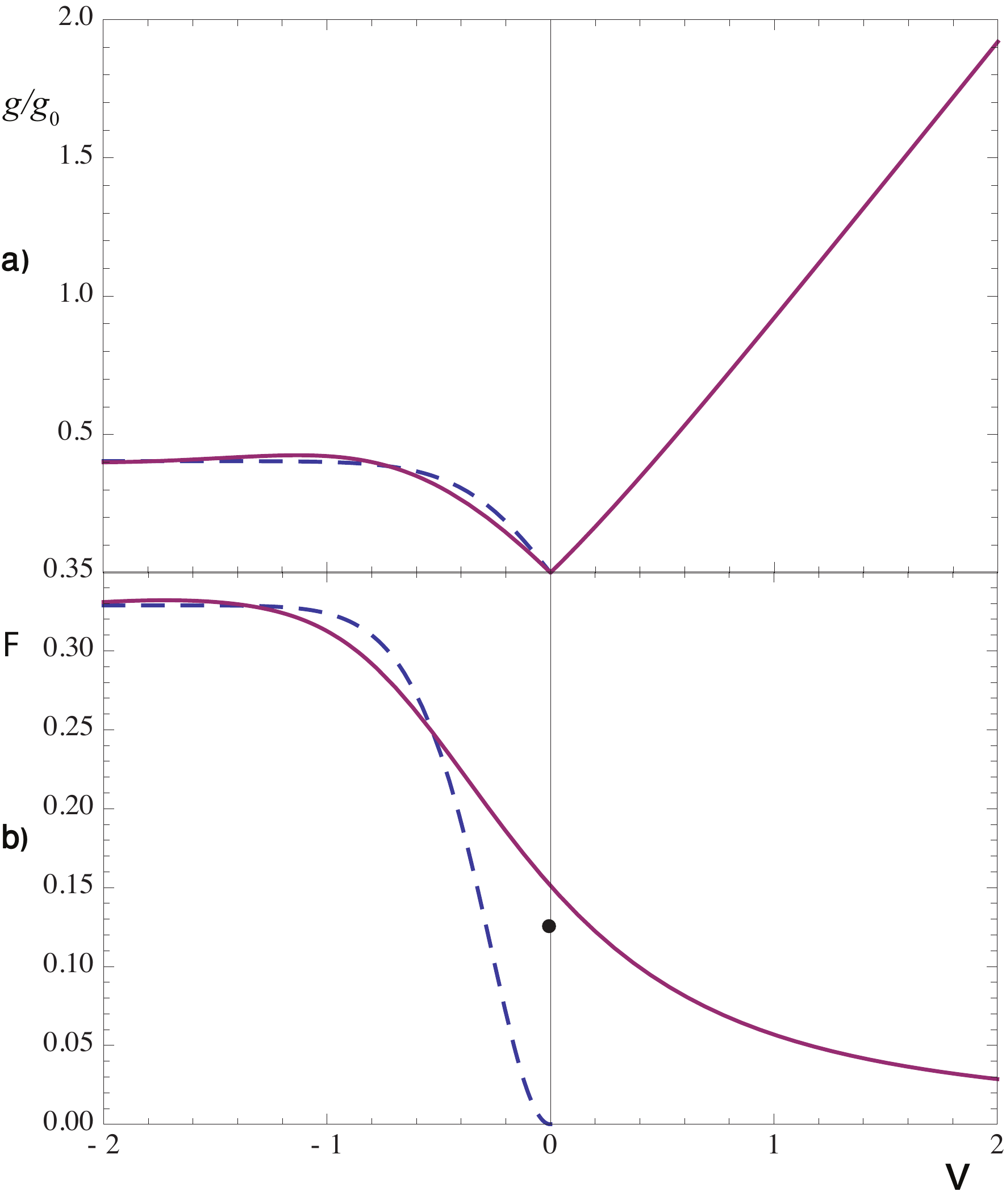}}
\caption{(color online) The plots (solid lines) reduced conductance $ g/g_0$ vs. gate voltage (a) and Fano factor $F$ vs. gate voltage (b) for incoherent transport.  The dashed lines show the results of simplified calculations approximating the transmission probability at negative $V_g$ with the probability of  the Klein tunneling in a uniform electric field. The small black circle in (b) shows the value 0.125 of the Fano factor  for vertical slopes (rectangular barrier) at voltages $V_g \gg \hbar v_F /eL$.} 
\label{fig2}
\end{figure}

Figure \ref{fig2}a shows the reduced conductance $ g/g_0$ as a function of the reduced gate voltage $v=eV_g/\sqrt{a}\hbar v_F $.  At positive $V_g$ the conductance grows roughly linearly, which is related with the linear growth  of the density of the states with the voltage. One may compare the conductance in Fig.~\ref{fig2}a  with the conductance for a steep potential step (rectangular barrier): 
\begin{equation}
g_r=g_0 \int_0^{e|V_g|/\hbar v_F}Tdk_y={\pi\over 4}g_0 {e|V_g|\over \hbar v_F}. 
\end{equation}
Here  $T=k_x/k$. The plot in Fig.~\ref{fig2}a coincides with this dependence at very small $v$. Figure \ref{fig2}b shows the dependence of the Fano factor,
\begin{equation}
F=\frac{  \int_0^{|v|}T(1-T) d\kappa}{  \int_0^{|v|}T d\kappa} , 
     \label{fano}      \end{equation}
on the reduced gate voltage $v=eV_g/\sqrt{a}\hbar v_F $. The small black circle shows the value 0.125 of the Fano factor  for vertical slopes (rectangular barrier)\cite{ES} at voltages $V_g \gg \hbar v_F /eL$.

\begin{figure}%[t]
\centerline{\includegraphics[width=0.7\linewidth]{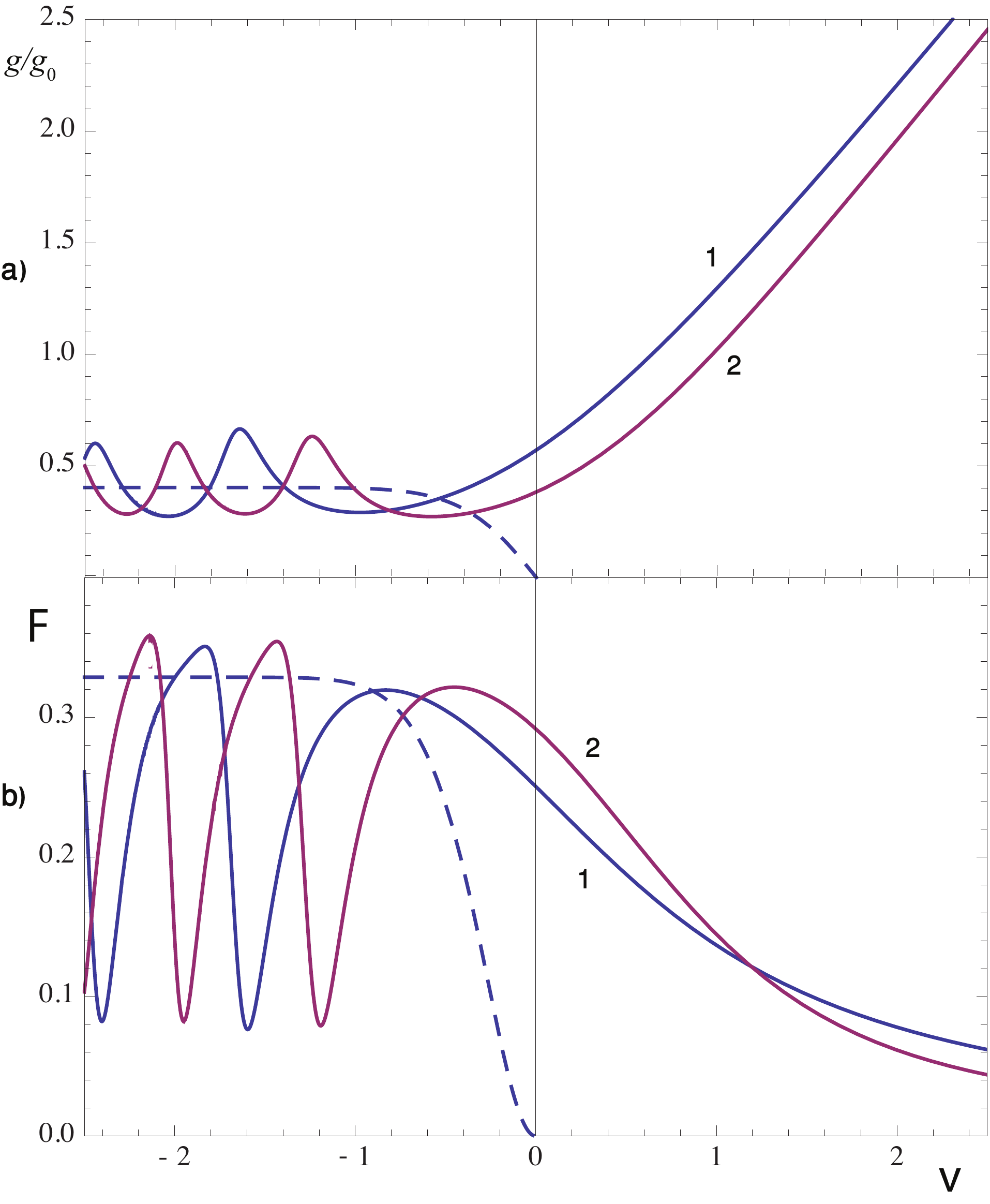}}
\caption{(color online) The plots  reduced conductance $ g/g_0$ vs. gate voltage (a) and Fano factor $F$ vs. gate voltage (b) for coherent transport.  Solid lines: 1 -- $L\sqrt{a}=0$, 2 -- $L\sqrt{a}=1$. The dashed lines show simplified calculations based on the probability of the Klein tunneling in a uniform electric field.  } 
\label{fig4}
\end{figure}

\begin{figure}%[t]
\centerline{\includegraphics[width=0.7\linewidth]{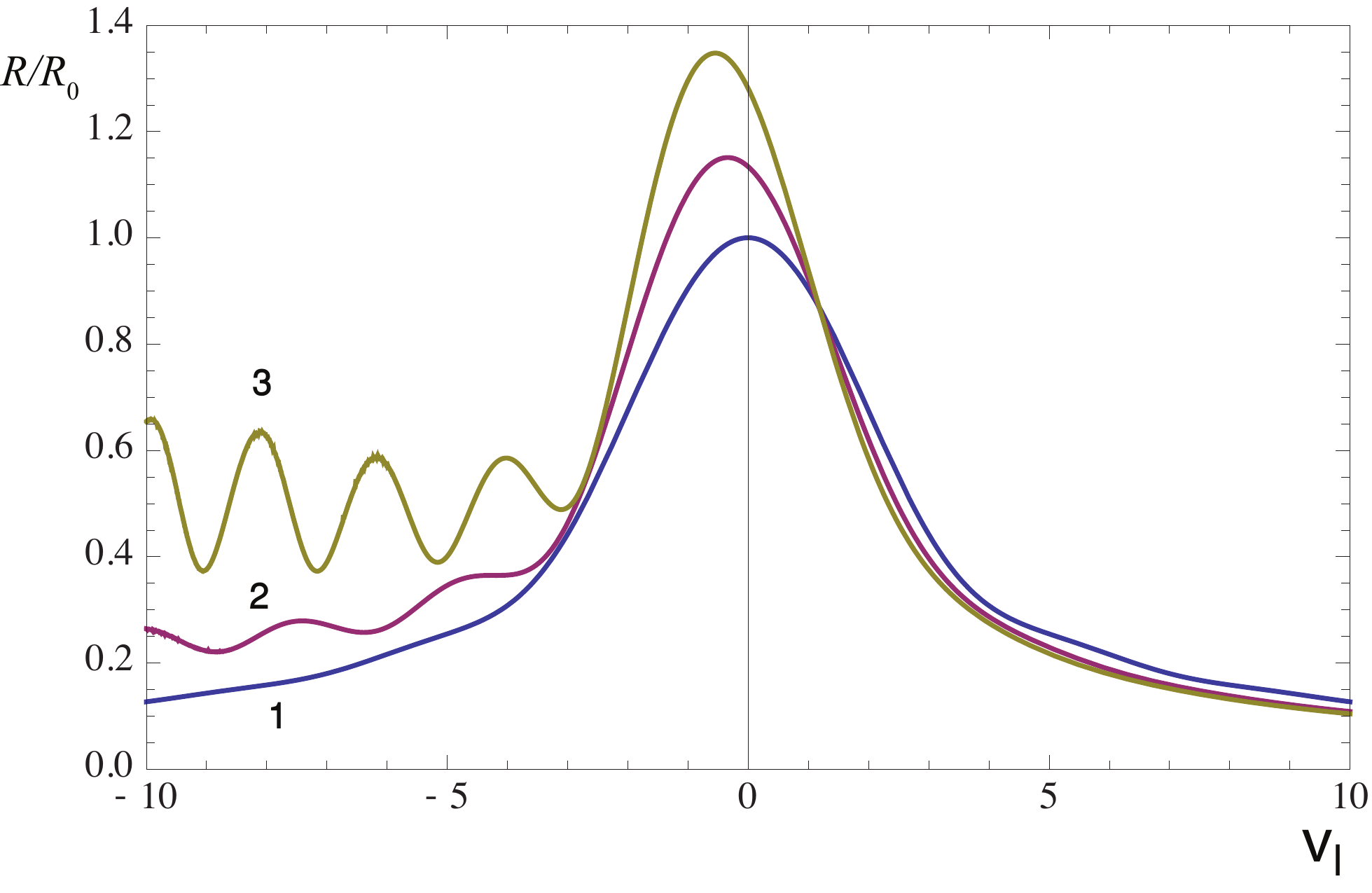}}
\caption{(color online) The plot resistance vs. gate voltage for coherent transport.  The lines 1, 2, and 3 correspond to the values $L\sqrt{a}=\infty, 10,$ and 5 respectively.  } 
\label{fig5}
\end{figure}

The left-hand parts of the plots (negative $V_g$) can be compared with the calculations assuming  that the transmission $T_1$ of a potential step is fully determined by the Klein-tunneling probability $T_K$ [see  Eq.~(\ref{KLZ})], and $T=T_K/(2-T_K)$ at any voltage. Then the conductance and the Fano factor are  
\begin{equation}
g=g_0  \int_0^{|v|}  {d \kappa \over 2e^{\pi \kappa^2}-1},~~~
F=\frac{  \int_0^{|v|}{ 2(e^{\pi \kappa^2}-1) d \kappa \over (2e^{\pi \kappa^2}-1)^2}}{  \int_0^{|v|}{d \kappa \over 2e^{\pi \kappa^2}-1}}.
       \end{equation}
They are  shown in Figs.~\ref{fig2}a and \ref{fig2}b by dashed lines. This approximation is similar to that used by Cheianov and Falko\cite{CF} for a single $p$ -- $n$ transition, but our analysis addresses two $p$ -- $n$ transitions in series.
One can see that at large negative gate voltage $V_g$ the conductance and the Fano factor are well described by the process of two sequential  uncorrelated Klein tunnelings and saturate at the plateaus determined by 
\begin{equation}
g_K=0.403 g_0,~~~F_K=0.329.
       \end{equation}
The subscript $K$ stresses that these values are determined by the Klein tunneling only, and their observation is a direct evidence of the Klein tunneling.

\begin{figure}%[t]
\centerline{\includegraphics[width=0.7\linewidth]{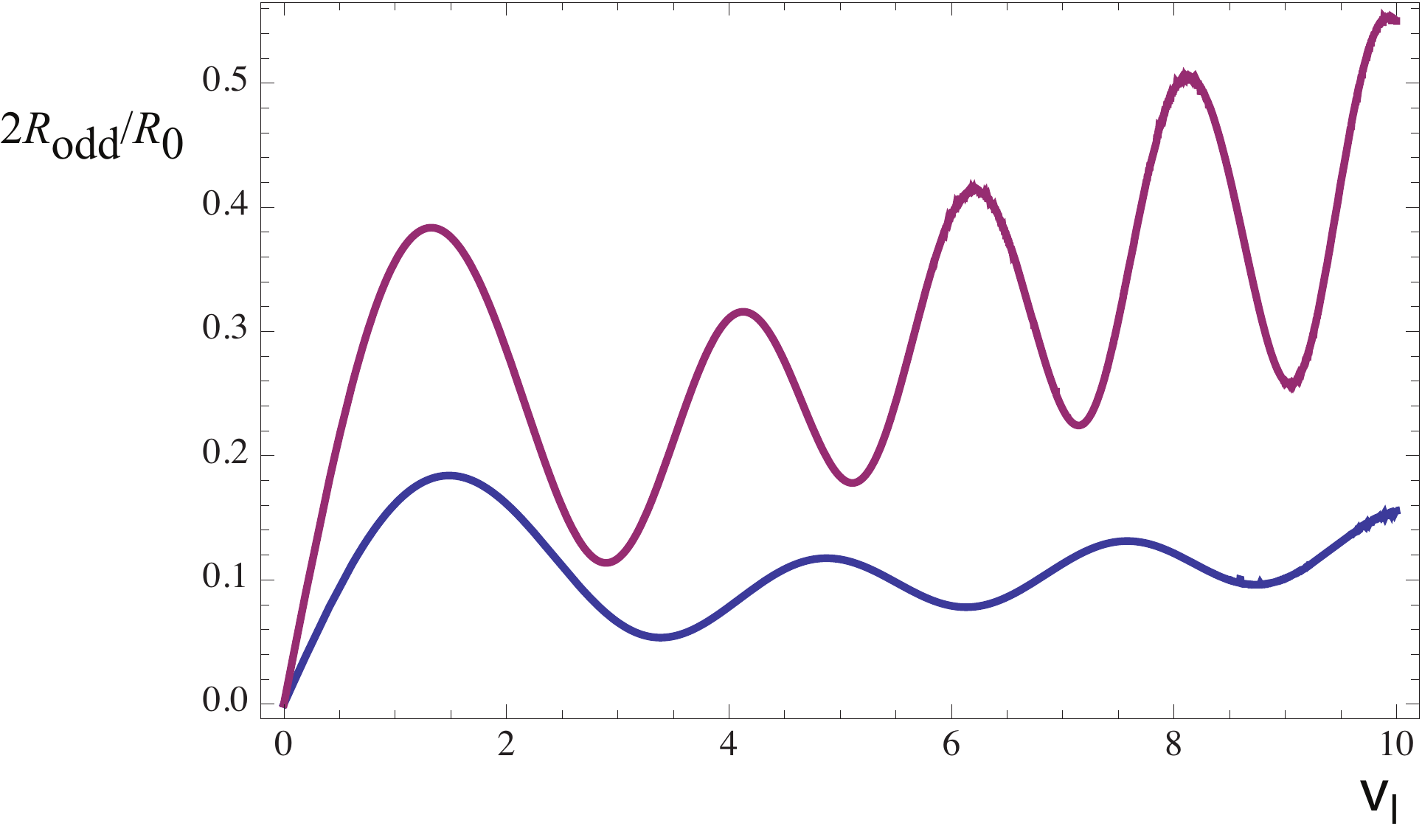}}
\caption{(color online) The plot odd resistance vs. gate voltage for coherent transport.  The lines 1 and  2 correspond to the values $L\sqrt{a}=$10  and 5 respectively. } 
\label{fig6}
\end{figure}

As mentioned above the content of this subsection refers to very large $L$. However this approach at some $L$ large enough can fail because of scattering by disorder. Thus the validness of the approach is restricted with the window $\hbar v_F/eV_g \ll L \ll l$ where $l$ is the mean-free path determined by disorder.

\subsection{Coherent ballistic transport}

In the coherent transport the total transmission amplitude and probability do depend on the length $L$ of the field-free region. For finite length $L$ the evanescent states participate in the transport, and in the expressions for the conductance and the Fano factor, Eqs.~(\ref{con}) and (\ref{fano}), one must replace the upper limit $|v|$ in the integrals by $\infty$.  Transmission $T=|t|^2$ in these expressions is now determined by Eq.~(\ref{totA}).
The numerically calculated conductance and Fano factor as functions of $v$ are shown for  $L\sqrt{a}=0$ and $L\sqrt{a}=1$  in Fig.~\ref{fig4}. In the case of  $L\sqrt{a}=0$ the barrier transforms from trapezoid to triangular.  As well as in Fig. \ref{fig2}, dashed lines show dependences  for the double Klein tunneling. One can see that  at negative $V_g$  the conductance and the Fano factor for the coherent transport oscillate around the ``Klein'' plateau.

The plots given above used the scales connected with the the length scale $1/\sqrt{a}$, which is determined by the slope only. This length is more useful until it less than $L$, i.e, for not too steep slopes.  However, in the opposite case of steep slope (a barrier close to rectangular) the length scale $L$ might be more useful. Then it would be convenient to use $G_0=1/R_0=4e^2 W /\pi h L$ (minimal conductance for a rectangular barrier) as a conductance scale and $\hbar v_F/eL$ as a voltage scale.  Figure~\ref{fig5}  shows the plots of the reduced resistance $R/R_0$ as a function of the dimensionless voltage $v_l =v L\sqrt{a} =eV_g L / \hbar v_F$ for three values of $L\sqrt{a}=\infty, 10,$ and 5.   Figure~\ref{fig6}  shows the dependence of the odd part of the resistance $R_{\mbox{odd}}=[R(-v)-R(v)]/2$ on the reduced voltage $v_l$ for two values of $L\sqrt{a}=10$ and 5 (at $L\sqrt{a}=\infty$ the dependence is symmetric and $R_{\mbox{odd}}$ vanishes). At large voltages $2 R_{\mbox{odd}}$ oscillates around the Klein resistance
\begin{equation}
R_K ={1\over g_K} = \frac{\pi h}{1.612 e^2 W \sqrt{a}}= \frac{\pi h\sqrt{d}}{1.612 e^2 W (\pi n_0)^{1/4}},
\label{RK} \end{equation}
where $n_0=k_0^2/\pi = eV_0 /\hbar v_F$ is the charger density in electrodes.

\section{Discussion and comparison with experiment} \label{last}

The odd part of the resistance was examined experimentally in Refs.~\onlinecite{Gold,Gold2,Gold1} (see Figs.~3 in Refs.~\onlinecite{Gold,Gold1} and Fig.~2 in Ref.~\onlinecite{Gold2}). There is some similarity between experimental curves and  theoretical curves in Fig.~\ref{fig6}:  the majority of experimental curves also have plateaus around which the curves oscillate. This supports the claim of Stander {\em et al.} that they have found evidence of the Klein tunneling. The oscillations in the experiment  look smaller and more chaotic than in the theory for coherent tunneling. This might be an effect of disorder. It is worthwhile to note that the appearance of Klein plateaus on theoretical curves is sensitive to the choice of the model: a plateau  can appear under the assumption that the electrical field inside the transient area of the slope does not vary and the gate voltage $V_g$ does not approach too close to the voltage $V_0$, which determines the potential step (Fig.~\ref{fig1}). The presence of plateaus on experimental curves  demonstrates that this assumption is not so bad. For further comparison we use the data for the sample  shown in Fig.~3a of Ref.~ \onlinecite{Gold1}.

Let us consider the dependence of the Klein resistance $R_K$ (resistance at the plateau) on the electric field. If the width $d$ of the transient area of the slope is fixed, the electric field is proportional to $V_0$, the latter being related with the charge density $n_0 $ in the electrodes. In the experiment  $n_0 $ changes from $1.2 \times 10^{12}$ cm$^{-2}$ to 
 $4.7 \times 10^{12}$ cm$^{-2}$. According to Eq.~(\ref{RK}) $R_K \propto  n_0^{-1/4}$ must decrease in 1.4 times while in the experiment the plateau resistance decreases in about 1.6 times.  A quantitative comparison of absolute values of the resistance is not straightforward because of the lack of information on the experimental value of the width $d$ of the slope area (the width of the $p$ -- $n$ transition). Choosing the distance between the top gate and the graphene sheet as a rough estimation for  $d$ (34 nm for the sample under consideration), for $n_0=4.7 \times 10^{12}$ cm$^{-2}$ Eq.~(\ref{RK}) yields resistance 0.110 k$\Omega$ against about 0.125  k$\Omega$ in the experiment. 

To conclude, the paper presents calculations of the conductance and the Fano factor in a graphene sheet in the ballistic regime. The electrostatic potential in the sheet is modeled by a trapezoid barrier, which allows to use the exact solution of the Dirac equation in a uniform electric field in the slope areas (the two lateral sides of the trapezoid). A special attention is devoted to asymmetry with respect to the sign of the gate voltage, which is connected with the difference between the Klein tunneling and the over-barrier reflection. The asymptotic resistance for large negative gate voltage, when an electron crosses  two $p$ -- $n$ transitions in series, is determined by the process of the Klein tunneling. The comparison of this asymptotic resistance (Klein resistance) with the experiment supports the conclusion that the Klein tunneling was revealed experimentally.

\section*{Acknowledgments}

I thank Pertti Hakonen for interesting discussions. The work was supported by the grant of the Israel Academy
of Sciences and Humanities.

\end{document}